\documentclass[preprint,aps,amssymb]{revtex4}

\usepackage{graphicx} 

\begin{document}

\title{Critical Phenomena in the Einstein-Massless-Dirac System}

\author{
Jason F.~Ventrella$^{1,2,3}$\footnote{{\tt ventrella@alum.mit.edu}}
\ and
Matthew W.~Choptuik$^{2,3,4}$\footnote{{\tt choptuik@physics.ubc.ca}}
}
\affiliation{
${}^{1}$Department of Physics and Astronomy, Louisiana State University, Baton Rouge, LA 70803-4001 USA\\
${}^{2}$Department of Physics and Astronomy, University of British Columbia,\\
Vancouver BC, V6T 1Z1 Canada\\
${}^{3}$Center for Relativity, University of Texas at Austin, TX 78712-1081 USA \\
${}^{4}$CIAR Cosmology and Gravity Program\\
}

\begin{abstract}
We investigate the general relativistic collapse of spherically 
symmetric, massless spin-$\frac{1}{2}$ fields at the threshold of black hole 
formation.  A spherically symmetric system is constructed from 
two spin-$\frac{1}{2}$ fields
by forming a spin singlet with no net spin-angular momentum.  We study the 
system numerically and find strong evidence for a Type II critical solution 
at the threshold between dispersal and black hole formation, with an associated 
mass scaling exponent $\gamma \sim 0.26$.  Although the critical solution 
is characterized by a continuously self-similar (CSS) geometry, the 
matter fields exhibit discrete self-similarity with an echoing exponent
$\Delta \sim 1.34$. 
We then adopt a CSS ansatz and reduce the 
equations of motion to a set of ODEs.
We find 
a solution of the ODEs that is analytic throughout the solution
domain, and show that it  corresponds to the critical solution found 
via dynamical evolutions.
\end{abstract}

\maketitle 

\section{Introduction}
\label{sec:intro}
Beginning with an investigation of the spherically symmetric collapse 
of a massless scalar field~\cite{crit}, many studies of gravitational 
collapse have established that the threshold of black hole formation 
is characterized by critical phenomena analogous to those that accompany
phase transitions in statistical mechanical systems.  Briefly,
(for a detailed review, see~\cite{gundlach}), black-hole threshold 
phenomena arise from the consideration of families of spacetimes, 
(generally containing one or more matter fields), which are labeled by a 
family parameter, $p$, that controls the degree of self-gravitation in the spacetime.
These families typically describe the implosion of an initially in-going 
concentration of matter-energy. For small values of $p$, the energy implodes 
in an essentially linear fashion, re-emerges and disperses to large distances.
In contrast, for large $p$, the implosion results in black hole formation, 
with some fraction of the initial mass of the system trapped within a horizon.
The threshold of black hole formation is defined by the specific (critical) 
parameter value, $p^{\star}$, at which a black hole first makes an appearance.

Empirically, and quite generically, a number of intriguing features are 
seen in the near-critical regime. These include the emergence of specific
critical solutions with scaling, or time-translational, symmetries, 
scaling laws of dimensionful quantities such as the black-hole mass, and 
universality in the sense that these features do not depend on the details 
of the family used to generate the critical solution.  There is now a relatively 
complete, though non-rigorous, understanding of this phenomenology.  First,
for a given matter model and symmetry restrictions (spherical, or axial symmetry 
for example), the black hole threshold apparently defines specific solutions 
of the coupled matter-Einstein equations.  These solutions are essentially 
unique, up to certain rescalings, or at least are isolated in the overall solution 
space of the model.  Second, these critical solutions, although unstable 
essentially by construction, tend to be minimally unstable in the sense of 
possessing a single (or perhaps a few) unstable modes in perturbation theory. 
The Lyapunov exponents associated with these modes can be immediately related 
to the exponents determined from empirically measured scaling laws.

In this paper we study the critical collapse of a massless spin-$\frac{1}{2}$ Dirac field 
coupled to the general relativistic gravitational field, within the context 
of spherical symmetry.  Since a single spinor field cannot be spherically 
symmetric, we consider the incoherent sum of two independent fields so that 
the superposition has no net spin-angular momentum. 
Through direct solution of the 1+1 dimensional 
PDEs governing the dynamics, we demonstrate the existence of so-called Type 
II behavior in the model, in which black hole formation turns on at infinitesimal 
mass, and the critical solution is self-similar.  In the current instance, 
the self-similarity
is somewhat novel in that individual components of the Dirac field are 
discretely self-similar, but the overall geometry is continuously self-similar.
As expected for this type of critical solution, we find a black hole mass-scaling law 
for solutions in the super-critical regime.
We then directly construct the threshold solution using an 
appropriate self-similar ansatz for the geometry and matter fields and 
demonstrate, good agreement between it and the PDE solution.

\section{Formalism}
\label{sec:form}
We consider a spherically symmetric system of spin-$\frac{1}{2}$ fields.
We use the ADM
formalism (see \cite{york} for details), adopt units in which 
$G=c=\hbar=1$, and express the metric in polar-areal 
coordinates:
\begin{equation}
\label{metric}
ds^{2} = - \alpha(t,r)^{2}dt^{2} + a(t,r)^{2}dr^{2} + r^{2}d\theta^{2} +
r^{2}\sin^{2}\theta d\phi^{2}.
\end{equation}
The coordinate $r$ is the {\it areal} radius defined as
$(A/4\pi)^{\frac{1}{2}}$, where $A$ is the proper area of a
constant-$r$ $2$-sphere.  The
functions $\alpha(t,r)$ and $a(t,r)$ are to be determined using 
a subset of the $3 + 1$ form
of Einstein's equations, as described in more detail 
in Section~\ref{sec:Geometry}.  

Before discussing how we separate the radial and
angular dependences in our system to form a spin singlet, we begin with a 
brief review of spinors in curved spacetime 
(see \cite{brill}, \cite{unruh}, and \cite{bd}).
The evolution of a massless, spin-$\frac{1}{2}$ field coupled to gravity 
is governed by the curved space Dirac equation:
\begin{equation}
\label{dirac}
\gamma^{\mu}\nabla_{\mu} \psi = 0.
\end{equation}
The curved space $\gamma$-matrices, $\gamma^{\mu}$ satisfy 
\begin{equation}
\label{clifford}
g^{\mu\nu}\openone = \frac{1}{2} \left\{ \gamma^{\mu},\gamma^{\nu} \right\}
\end{equation}
where $\openone$ is the $4 \times 4$ identity matrix and $g^{\mu\nu}$ is the 
inverse metric.
In flat spacetime, we have:
\begin{equation}
\label{clifflat}
\eta^{ab}\openone = \frac{1}{2} \left\{ \tilde \gamma^{a},\tilde
\gamma^{b}
\right\}.
\end{equation}
A particular choice of $\tilde \gamma^{a}$ that satisfy (\ref{clifflat})
with our metric signature $(-,+,+,+)$ is:
\begin{equation}
\label{flatgamma}
\tilde \gamma^{0} = i \left(
\begin{array}{cc}
              \openone & 0  \\
              0 & - \openone
\end{array}
\right),
\hspace{0.5in}
\tilde \gamma^{j} = i \left(
\begin{array}{cc}
              0 & \sigma^{j}  \\
              - \sigma^{j} & 0
\end{array}
\right).
\end{equation}
Here, the index $j$ ranges over the spatial values 
$1,2,3$ and the $\sigma^{j}$ are the Pauli spin matrices, namely
\begin{equation}
\sigma^{1} = \left(
\begin{array}{cc}
              0 & 1  \\
              1 & 0
\end{array}
\right),
\sigma^{2} = \left(
\begin{array}{cc}
              0 & -i  \\
              i & 0
\end{array}
\right),
\sigma^{3} = \left(
\begin{array}{cc}
              1 & 0  \\
              0 & -1
\end{array}
\right).
\end{equation}
 The general $\gamma$-matrices are related to their flat,
Cartesian counterparts by
\begin{equation}
\label{rep}
\gamma^{\mu} = {V_a{}^\mu}\tilde \gamma^{a}
\end{equation}
where there is an implied summation over the values 
$0,1,2,3$ of the ``flat'', Latin index $a$, and the
${V_a{}^\mu}$ are known as vierbein.

The derivative operator in equation (\ref{dirac}) is a {\it spinor} covariant 
derivative with spinor affine connections, $\Gamma_{\mu}$.  It acts in the 
following way on spinors,
\begin{equation}
\nabla_{\mu} \psi = \left( \frac{\partial}{\partial x^{\mu}} - \Gamma_{\mu}
\right)\psi,
\end{equation}
and on $\gamma$-matrices,
\begin{equation}
\nabla_{\mu} \gamma^{\nu} = \frac{\partial}{\partial x^{\mu}} \gamma^{\nu} 
+ {\Gamma^\nu{}_{\mu\lambda}} \gamma^{\lambda} - \Gamma_{\mu}\gamma^{\nu} 
+ \gamma^{\nu} \Gamma_{\mu}.
\end{equation}
However, it reduces to the usual covariant derivative when acting on tensors.  
We choose the spinor connections so that
\begin{equation}
\nabla_{\mu} \gamma^{\nu} = 0.
\end{equation}
It can then be shown that the $\Gamma_\mu$ take the form
\begin{equation}
\Gamma_{\mu} =-\frac{1}{8}\left[\tilde \gamma^{a},\tilde \gamma^{b}\right]
{V_a{}^\nu}\nabla_{\mu}V_{b\nu}.
\end{equation}
We also note that 
when taking the covariant derivative of the vierbein, $V_a{}^\mu$, 
only one Christoffel connection appears, since there is only one 
curved, tensor (Greek) index.

\subsection{Representation}
Having fixed the form of the spherically symmetric metric 
(\ref{metric}), we are ready to 
find a set of $\gamma$-matrices that satisfy equation (\ref{clifford}).
We choose as our representation:
\[
\gamma^{t} = \frac{\tilde \gamma^{0}}{\alpha},\hspace{0.3in}
\gamma^{r} = \frac{\tilde \gamma^{3}}{a},
\]
\begin{equation}
\label{myrep}
\gamma^{\theta} = \frac{\tilde \gamma^{2}}{r},\hspace{0.1in}
\gamma^{\phi} = \frac{\tilde \gamma^{1}}{r \sin\theta}.
\end{equation}
The spinor connections are
\[
\Gamma_{t} = \frac{1}{2} \frac{\alpha^{\prime}}{a} 
\tilde \gamma^{0} \tilde \gamma^{3},
\]
\[
\Gamma_{r} = \frac{1}{2} \frac{\dot{a}}{\alpha}
\tilde \gamma^{0} \tilde \gamma^{3},
\]
\[
\Gamma_{\theta} = \frac{1}{2} \frac{1}{a}
\tilde \gamma^{3} \tilde \gamma^{2},
\]
\begin{equation}
\Gamma_{\phi} = \frac{1}{2} \frac{\sin\theta}{a}\tilde \gamma^{3}
\tilde \gamma^{1} + \frac{1}{2}\cos\theta \tilde \gamma^{2} \tilde \gamma^{1}.
\end{equation}
where dots and primes denote differentiation with respect to $t$ and $r$
respectively. 
We note that we have complete freedom to choose any set of
$\gamma^{\mu}$ we wish provided they satisfy equation (\ref{clifford}),
and that
our specific choice is made so that the Dirac equation can be easily
separated into radial and angular parts.

Before proceeding to this separation, we introduce a further simplification
based on the fact 
that we are dealing with a massless spin-$\frac{1}{2}$ field.  
Mathematically, such a field has a particular chirality (circular
polarization); we adopt left-handed chirality which is expressed as:
\begin{equation}
\label{chiral}
\left( \openone - i \gamma^{5} \right)\psi = 0.
\end{equation}
Here, $\gamma^{5}$ is defined by
\begin{equation}
\gamma^{5} \equiv \tilde \gamma^{0} \tilde \gamma^{1} \tilde \gamma^{2} \tilde
\gamma^{3}.
\end{equation}
Equation (\ref{chiral}) can be satisfied by taking
\begin{equation}
\psi = \left( 
\begin{array}{c}
\psi_{1}(t,r,\theta,\phi)\\
\psi_{2}(t,r,\theta,\phi)\\
\psi_{1}(t,r,\theta,\phi)\\
\psi_{2}(t,r,\theta,\phi)
\end{array}
\right).
\end{equation}
Substitution of this form for the spinor into equation (\ref{dirac}) 
yields two identical sets 
of equations, each coupling the spinor components, $\psi_{1}$ and $\psi_{2}$.  
We of course only need to solve one set of equations for these variables, so 
we are left with two equations instead of the original four.

We now perform a separation of variables on the spinor components by
writing: 
\begin{equation}
\left(
\begin{array}{c}
\psi_{1}(t,r,\theta,\phi)\\
\psi_{2}(t,r,\theta,\phi)
\end{array}
\right) = \frac{1}{r \sqrt{a(t,r)}}
\left(
\begin{array}{c}
\label{split}
F(t,r)H_{1}(\theta,\phi)\\
G(t,r)H_{2}(\theta,\phi)
\end{array}
\right).
\end{equation}
With this new choice of variables, and with our previously chosen 
representation of the $\gamma^\mu$~(\ref{myrep}), the Dirac equation 
separates into a part that depends
on $(t,r)$ (which we will refer to as the ``radial'' part), and a 
part that depends on $(\theta,\phi)$:
\begin{eqnarray}
\label{sep}
\frac{ir}{\alpha} \left( 
\begin{array}{c}
\dot{F}/G \\
\dot{G}/F 
\end{array}
\right) + \frac{ir}{2} \frac{a^{\prime}}{a^{2}} \left(
\begin{array}{c}
- F/G \\
G/F
\end{array}     
\right)
+ \frac{ir}{a} \left(
\begin{array}{c}
F^{\prime}/G \\
G^{\prime}/F      
\end{array}             
\right)
+ \frac{ir}{2}\frac{\alpha^{\prime}}{\alpha a} \left(
\begin{array}{c}
F/G \\
- G/F
\end{array}
\right) \nonumber\\
+ \frac{i}{\sin\theta}
\left(
\begin{array}{c} 
H_{2,\phi}/H_{1} \\
H_{1,\phi}/H_{2}
\end{array}                             
\right) 
+ \left(
\begin{array}{c}
H_{2,\theta}/H_{1} \\
- H_{1,\theta}/H_{2}
\end{array}               
\right) 
+ \frac{1}{2}\cot\theta \left(
\begin{array}{c}
H_{2}/H_{1} \\
- H_{1}/H_{2}
\end{array}                      
\right) = 0.
\end{eqnarray}
We note that although the factor $(r\sqrt{a})^{-1}$ in~(\ref{split})
is not necessary for the separation of variables, it simplifies matters by 
reducing the number of terms in the resulting equation.
Of particular importance is 
the elimination of a time derivative of $a(t,r)$ that would make numerical
solution of the resulting system of PDEs somewhat more involved.

Considering our separated equation, we observe that 
since any change in $\theta$ 
or $\phi$ cannot change the value of 
the $(t,r)$ part of (\ref{sep}), the $(\theta,\phi)$ part must be a 
constant.  At this point, if our goal was to 
simply remove the angular dependence from the Dirac equation, we would be done.
By replacing the angular part of (\ref{sep}) by a constant we would be 
restricting ourselves to some spinor that is an eigenfunction of the angular 
operators in the Dirac equation.  In fact only one of its eigenvalues, 
rather than the precise form of the angular eigenfunction,
would need to be known.
However, our goal is not only to eliminate the angular dependence of our 
equation of motion, but also to have a matter source that generates a 
spherically symmetric spacetime.  An individual spinor is {\it not} a spherically 
symmetric object (it always has a spin-angular momentum that breaks this 
symmetry) and therefore cannot by itself produce such a spacetime.  What we 
require are multiple spinors where all the individual 
spin-angular momenta counterbalance each other so the system has no net spin.
We will use two spinors, for simplicity, but any even number of the appropriate
spinors could also be used.
The spherically symmetric stress-energy tensor for the
system, $T_{\mu\nu}$, is found from the sum of the stress-energy tensors of the
individual spinor fields \cite{bill}. 
\begin{equation}
\label{incsum}
T_{\mu\nu} = T^{+}_{\mu\nu} + T^{-}_{\mu\nu} 
\end{equation}
Evaluating the right hand side of equation
(\ref{incsum}) {\it does} require the angular eigenfunctions that we will now 
compute.

\subsection{Equations of Motion}
Setting the angular part of equation (\ref{sep}) equal to a constant, $n$, 
gives:
\begin{equation} 
\label{eth}
\left[ - \frac{i}{\sin\theta}
\frac{\partial }{\partial \phi} - \frac{\partial }{\partial \theta}
- \frac{1}{2}\cot\theta \right] H_{2} = - n H_{1},
\end{equation} 
\begin{equation} 
\label{ethbar}
\left[ \frac{i}{\sin\theta}
\frac{\partial }{\partial \phi} - \frac{\partial }{\partial \theta}
- \frac{1}{2}\cot\theta \right] H_{1} = n H_{2},
\end{equation} 
where we have multiplied the first and second components of the 
angular terms in~(\ref{split}) by $-H_1$ and $H_2$ respectively,
so that the bracketed terms in the above expression are the raising and 
lowering operators, $\eth$ (eth) and $\bar{\eth}$ (ethbar), respectively.  
These operators 
act on the spin weighted spherical harmonics, 
$\mathord{{}_{s\mkern-4mu}Y_{lm}}$,
(see \cite{penrose} and \cite{goldberg})
in the following way:
\begin{equation}
\label{raise}
\eth(\mathord{{}_{s\mkern-4mu}Y_{lm}}) = \sqrt{(l-s)(l+s+1)}
(\mathord{{}_{s+1\mkern-4mu}Y_{lm}})
\end{equation}
\begin{equation}
\label{lower}
\bar{\eth}(\mathord{{}_{s\mkern-4mu}Y_{lm}}) = - \sqrt{(l+s)(l-s+1)}
(\mathord{{}_{s-1\mkern-4mu}Y_{lm}}).
\end{equation}
Our functions $H_{1}$ and $H_{2}$ have the spin weights $s = \pm \frac{1}{2}$:
\begin{equation}
H_{1}(\theta,\phi) = \mathord{{}_{\frac{1}{2}\mkern-4mu}Y_{lm}}(\theta,\phi)
\end{equation}
\begin{equation}
H_{2}(\theta,\phi) = \mathord{{}_{-\frac{1}{2}\mkern-4mu}Y_{lm}}(\theta,\phi).
\end{equation}

To form a spin singlet, all we require is one 
spinor constructed from 
$\mathord{{}_{\frac{1}{2}\mkern-4mu}Y_{\frac{1}{2}\frac{1}{2}}}$,
$\mathord{{}_{-\frac{1}{2}\mkern-4mu}Y_{\frac{1}{2} \frac{1}{2}}}$, and one 
spinor from $\mathord{{}_{\frac{1}{2}\mkern-4mu}Y_{\frac{1}{2}-\frac{1}{2}}}$,
$\mathord{{}_{-\frac{1}{2}\mkern-4mu}Y_{\frac{1}{2}-\frac{1}{2}}}$.  
These spin weighted spherical harmonics are:
\[
\mathord{{}_{\frac{1}{2}\mkern-4mu}Y_{\frac{1}{2}\frac{1}{2}}} = 
                   \frac{1}{\sqrt{ 2\pi}} e^{i\phi/2} \sin\frac{\theta}{2},
\]
\[
\mathord{{}_{-\frac{1}{2}\mkern-4mu}Y_{\frac{1}{2}\frac{1}{2}}} = 
                   \frac{1}{\sqrt{ 2\pi}} e^{i\phi/2} \cos\frac{\theta}{2},
\]
\[
\mathord{{}_{\frac{1}{2}\mkern-4mu}Y_{\frac{1}{2} -\frac{1}{2}}} = 
                   \frac{1}{\sqrt{ 2\pi}} e^{-i\phi/2} \cos\frac{\theta}{2},
\]
\begin{equation}
\mathord{{}_{-\frac{1}{2}\mkern-4mu}Y_{\frac{1}{2}-\frac{1}{2}}} = 
                    - \frac{1}{\sqrt{ 2\pi}} e^{- i\phi/2} \sin\frac{\theta}{2}
\end{equation}
and are solutions to (\ref{eth}) and (\ref{ethbar}) for $n = -1$. 
Thus, for our two spinor fields, we have:
\begin{equation}
\label{spinor+}
\psi^+ = \frac{1}{2\sqrt{\pi}}\frac{e^{i\phi/2}}{r\sqrt{a(t,r)}}\left(
\begin{array}{c}
F(t,r)\sin(\theta/2)\\
G(t,r)\cos(\theta/2)\\
F(t,r)\sin(\theta/2)\\
G(t,r)\cos(\theta/2)
\end{array}
\right)
\end{equation}
\begin{equation}
\label{spinor-}
\psi^- = \frac{1}{2\sqrt{\pi}}\frac{e^{-i\phi/2}}{r\sqrt{a(t,r)}}\left(
\begin{array}{c}
  F(t,r)\cos(\theta/2)\\
- G(t,r)\sin(\theta/2)\\
  F(t,r)\cos(\theta/2)\\
- G(t,r)\sin(\theta/2)
\end{array}
\right).
\end{equation}
From the above results, we can now derive the following radial equations 
of motion from~(\ref{sep}):
\begin{equation}
\label{eom1}
\dot{F}_{1} = - \sqrt{\frac{\alpha}{a}}\partial_{r}\left(
\sqrt{\frac{\alpha}{a}}F_{1}\right)  + \frac{\alpha}{r}G_{2}
\end{equation}
\begin{equation}
\dot{G}_{1} = \sqrt{\frac{\alpha}{a}}\partial_{r}\left(
\sqrt{\frac{\alpha}{a}}G_{1}\right) + \frac{\alpha}{r}F_{2} 
\end{equation}
\begin{equation}
\dot{F}_{2} = - \sqrt{\frac{\alpha}{a}}\partial_{r}\left(
\sqrt{\frac{\alpha}{a}}F_{2}\right) - \frac{\alpha}{r}G_{1}
\end{equation}
\begin{equation}
\label{eom4}
\dot{G}_{2} = \sqrt{\frac{\alpha}{a}}\partial_{r}\left(
\sqrt{\frac{\alpha}{a}}G_{2}\right) - \frac{\alpha}{r}F_{1} 
\end{equation}
where we have written the complex functions, $F(t,r)$ and $G(t,r)$ 
in terms of real functions via 
$F(t,r) \equiv F_{1}(t,r) + iF_{2}(t,r)$ and 
$G(t,r) \equiv G_{1}(t,r) + iG_{2}(t,r)$.

\subsection{Geometry}
\label{sec:Geometry}
Although the spinors (\ref{spinor+}) and (\ref{spinor-}) both yield the 
same radial equations of motion (\ref{eom1})-(\ref{eom4}), they have
different stress-energy tensors.  We calculate the stress-tensor for each
field individually using:
\begin{equation}
\label{stress}
T_{\mu \nu} = -\frac{1}{2}\left[ \bar{\psi}\gamma_{(\mu}\nabla_{\nu)}\psi
               -\left( \nabla_{(\mu}\bar{\psi}\right)\gamma_{\nu)}\psi \right]
\end{equation}
where the Dirac adjoint of $\psi, \bar{\psi}$ is defined by
\[
\bar{\psi} = \psi^{\dagger}A.
\]
Here, $A$ is the so-called Hermitizing matrix, needed in the computation of
real-valued expressions (such as the current density or, in this case, the
stress-energy tensor) from the complex-valued spinors.
It is to be chosen so that both $A$ and $iA\gamma^{\mu}$ are Hermitian, and 
we take $A = -i\tilde{\gamma}^{0}$.

Computing $T_{\mu\nu}$ for each of the two spinors 
(\ref{spinor+}) and (\ref{spinor-}) and 
summing the results
yields the following non-vanishing components of the 
spherically symmetric stress-energy tensor:
\[
T_{tt} = \frac{\alpha}{2\pi r^{2} a}\left( \dot{F}_{1}F_{2} - F_{1}\dot{F}_{2}
+ \dot{G}_{1}G_{2} - G_{1}\dot{G}_{2} \right)
\]
\[
T_{tr} = \frac{1}{4\pi r^{2}}\left[ F_{1}\dot{F}_{2} - \dot{F}_{1}F_{2}
+ \dot{G}_{1}G_{2} - G_{1}\dot{G}_{2} 
+ \frac{\alpha}{a} \left( F^{\prime}_{1}F_{2} - F_{1}F^{\prime}_{2}
+ G^{\prime}_{1}G_{2} - G_{1}G^{\prime}_{2} \right) \right]
\]
\[
T_{rr} = \frac{1}{2\pi r^{2}}\left( F_{1}F^{\prime}_{2} - F^{\prime}_{1}F_{2}
+ G^{\prime}_{1}G_{2} - G_{1}G^{\prime}_{2} \right)
\]
\[
T_{\theta \theta} = \frac{1}{2\pi r a} \left( F_{1}G_{1} + F_{2}G_{2} \right)
\]
\begin{equation}
\label{totstress}
T_{\phi \phi} = \frac{\sin^{2}\theta}{2\pi r a} \left( F_{1}G_{1} + F_{2}G_{2} 
\right).
\end{equation}
Contracting the stress-energy tensor gives
\begin{equation}
\label{cinv}
{T_\mu{}^\mu} = 0,
\end{equation}
which is expected since the massless Dirac system is conformally invariant.
Having computed a stress-energy tensor that will generate a spherically
symmetric spacetime, we can now write down the Einstein equations
that will fix $\alpha(t,r)$ and $a(t,r)$.

Due to our choice of coordinates, 
the Hamiltonian constraint and the
slicing condition comprise a sufficient subset of the Einstein equations to 
be used in our numerical solution.
The Hamiltonian constraint is
\begin{equation}
\label{hamcon}
\frac{a^{\prime}}{a} + \frac{a^{2} - 1}{2 r} = \frac{2}{r^{2}}
\left( 2 a F_{1}G_{1} + 2 a F_{2}G_{2} + rF_{1}F^{\prime}_{2} 
- rF^{\prime}_{1}F_{2} + rG^{\prime}_{1}G_{2} - rG_{1}G^{\prime}_{2} \right)
\end{equation}
and is treated as an equation for $a(t,r)$.  We note that the
momentum constraint,
\begin{equation}
\label{momcon}
\dot{a} = \frac{2\alpha}{r}\left( F^{\prime}_{1}F_{2} - F_{1}F^{\prime}_{2}
+ G^{\prime}_{1}G_{2} - G_{1}G^{\prime}_{2} \right).
\end{equation}
also yields an equation for $a$ (an evolution equation) that we use
as a means to check the consistency of our equations, both at the 
analytic level, and during numerical evolutions. 
We note that in both equations (\ref{hamcon}) and (\ref{momcon}), 
time derivatives of $F$'s
and $G$'s have been eliminated using the equations of motion
(\ref{eom1})-(\ref{eom4}).

The slicing condition, which fixes $\alpha(t,r)$, is derived from the
evolution equation for ${K^\theta{}_{\theta}}$ and the fact that for polar 
slicing we have
\[
K = {K^i{}_{i}} = {K^r{}_{r}} + 2 {K^\theta{}_{\theta}} = {K^r{}_{r}}
\]
since ${K^\theta{}_{\theta}}(t,r) = 0$.  
To maintain ${K^\theta{}_{\theta}}(t,r) =
0$ for all time, we impose ${\dot{K}^\theta{}_{\theta}}(t,r) = 0$,
which yields
\begin{equation}
\label{lapse}
\frac{\alpha^{\prime}}{\alpha} - \frac{a^{2}-1}{2r}
= \frac{2}{r}\left( F_{1}F^{\prime}_{2} - F^{\prime}_{1}F_{2}
+ G^{\prime}_{1}G_{2} - G_{1}G^{\prime}_{2} \right).
\end{equation}

\section{Numerics and Results}
\label{sec:res}

The above equations of motion were solved using a Crank-Nicholson 
update scheme, standard $O(h^{2})$ spatial derivatives, and Berger-Oliger 
style adaptive mesh refinement (see \cite{bo}) on the computational
domain, $0 \le r \le r_{\rm max}$, $t \ge 0$.  To achieve 
stability, high frequency modes were damped using Kreiss-Oliger 
dissipation \cite{ko}.  
At $r=0$, the following regularity conditions were enforced:
\begin{equation}
\begin{array}{rcl}
F_{1}(t,0) & = & 0 \\
F_{2}(t,0) & = & 0 \\
G_{1}(t,0) & = & 0 \\
G_{2}(t,0) & = & 0. \\
\end{array}
\end{equation}
At $r = r_{\rm max}$, outgoing wave conditions are imposed on the matter 
fields:
\begin{equation}
\begin{array}{rcl}
\partial_{t}F_{1} & = & -\partial_{r}F_{1}, \\
\partial_{t}F_{2} & = & -\partial_{r}F_{2}, \\
\partial_{t}G_{1} & = & -\partial_{r}G_{1}, \\
\partial_{t}G_{2} & = & -\partial_{r}G_{2}.\\
\end{array}
\end{equation}

We also have 
\begin{equation}
a(t,0) = 1,
\end{equation}
which follows from the demand of regularity (local flatness) at $r = 0$.
At each time step, the Hamiltonian constraint is integrated outwards from
$r=0$ using 
a point-wise Newton method. The slicing condition~(\ref{lapse}) is solved 
subject to the outer boundary condition 
$\alpha(t,r_{\rm max}) = 1/a(t,r_{\rm max})$, so that, provided 
that all of the matter remains within the computational domain, coordinate time 
and proper time coincide at spatial infinity.  We note that the more natural
normalization choice, from the point of view of critical collapse, is 
$\alpha(t,0) = 1$. However, computationally, this choice 
would vitiate our ability to compute with a fixed Courant factor,
$\Delta t / \Delta r$, particularly in the near-critical regime, and would 
thus unnecessarily complicate the numerical solution. 

The results presented below were generated from the study of three
distinct parametrized families of initial data, which we refer to as 
Gaussians, spatial derivatives of Gaussians, and kink--anti-kink.  
In all cases, the initial datasets were such that the pulses were 
essentially ingoing with positive energy at $t=0$.  The specific forms used 
are as follows:

{\it Gaussian}:
\[
F_{1} = 0
\]
\[
F_{2} = 0
\]
\[
G_{1} = p \, e^{-(r-r_{0})^{2}/4*\delta^{2}}
\]
\[
G_{2} = p \, e^{-(r-r_{0}+\delta)^{2}/4*\delta^{2}}
\]

{\it Derivative of Gaussian}:
\[
F_{1} = 0
\]
\[
F_{2} = 0
\]
\[
G_{1} = p \, ((r_{0}-r)/(2\delta^{2}))e^{-(r-r_{0})^{2}/4\delta^{2}}
\]
\[
G_{2} = p \, ((r_{0}-r-\delta)/(2\delta^{2}))e^{-(r-r_{0}+\delta)^{2}/4\delta^{2}}
\]

{\it Kink anti-Kink}:
\[
F_{1} = 0
\]
\[
F_{2} = 0
\]
\[
G_{1} = (p/2)(\tanh((r-r_{0})/\delta)-\tanh((r-2r_{0})/\delta))
\]
\[
G_{2} = (p/2)(\tanh((r-r_{0} + \delta)/\delta)-\tanh((r-2r_{0} + \delta)/\delta))
\]

In each case the family parameter, $p$, can be used to 
control whether 
the mass-energy of the system collapses to form a black hole, 
or if it 
implodes through the center ($r = 0$) and disperses to spatial infinity.  
As discussed in the introduction, and in a process now familiar from 
many other studies of critical collapse, 
as $p$ is tuned to the black hole threshold $p=p^{\star}$, the single 
unstable mode associated with the critical solution is ``tuned away'' 
to reveal the critical solution {\em per se}.

In the current case, we find strong indications from such studies that the 
critical solution for the spherically symmetric EMD (Einstein-massless-Dirac)
system describes a continuously self-similar (CSS) geometry.  Typical 
evidence for this claim is shown in  Fig.~\ref{aplot}, which displays 
near-critical 
evolution of the scale invariant quantity, $a$.
In contrast, the Dirac fields themselves appear to be discretely 
self-similar (DSS) (see Fig.~\ref{F1plot}).

As usual, associated with the Type II critical solution  is a scaling 
law for the black hole mass near criticality:
\begin{equation}
\label{bhscale}
M_{bh} \propto |p-p^{\star}|^{\gamma},
\end{equation}
and, as expected, we find strong evidence that the 
scaling exponent, $\gamma$, is 
universal in that it is independent of the family of initial data used to 
generate the critical solution.  The
values of $\gamma$ computed for the different families are summarized in
Table~\ref{scalexp}, and we note that there
is uncertainty in the third digit of the quoted values.
\begin{table}[h]
\begin{center}
\caption{\label{scalexp}Scaling exponent, $\gamma$, associated with the three families of initial data
described in the text.}
\vskip 10pt
\begin{tabular}{|l|l|}
\cline{1-2}
{\textsl{Family}} &
{\textsl{$\gamma$}} \\
\cline{1-2}
Gaussian               & 0.258 \\ 
derivative of Gaussian & 0.259 \\ 
kink anti-kink         & 0.257 \\ 
\cline{1-2}
\end{tabular}
\end{center}
\end{table}

\begin{figure}[ht]
\includegraphics[scale=0.5]{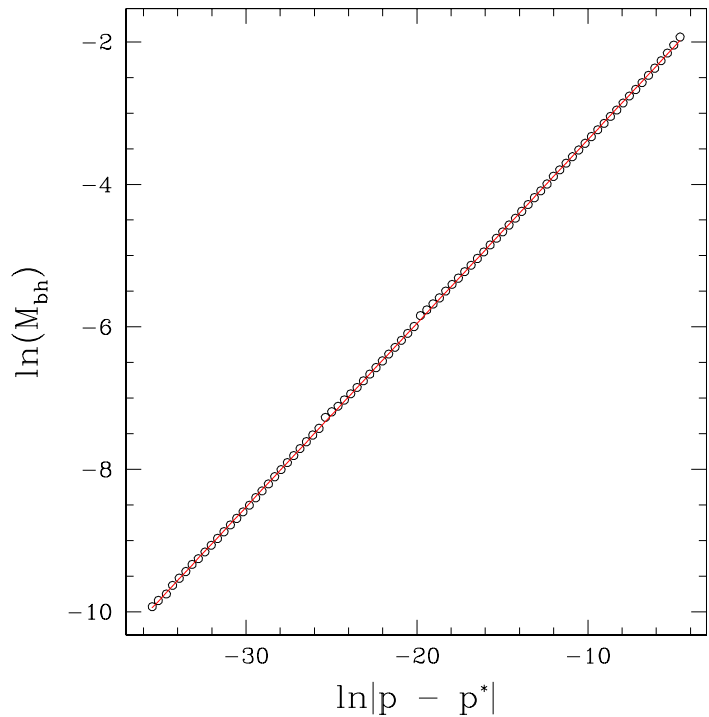}
\caption{Plot of the observed mass scaling near criticality for the 
case of the Gaussian family.
The measured scaling exponent 
is $\gamma = 0.258$, 
with uncertainty in the third digit.
As we
tune arbitrarily close to the critical parameter, $p^{\star}$, black holes
of arbitrarily small masses are formed, indicative of a Type II critical 
solution. The small irregularities visible in the plot are shown and 
discussed in more detail in Fig.~\ref{bherr}} 
\label{bhplot}
\end{figure}

\begin{figure}[ht]
\includegraphics[scale=0.5]{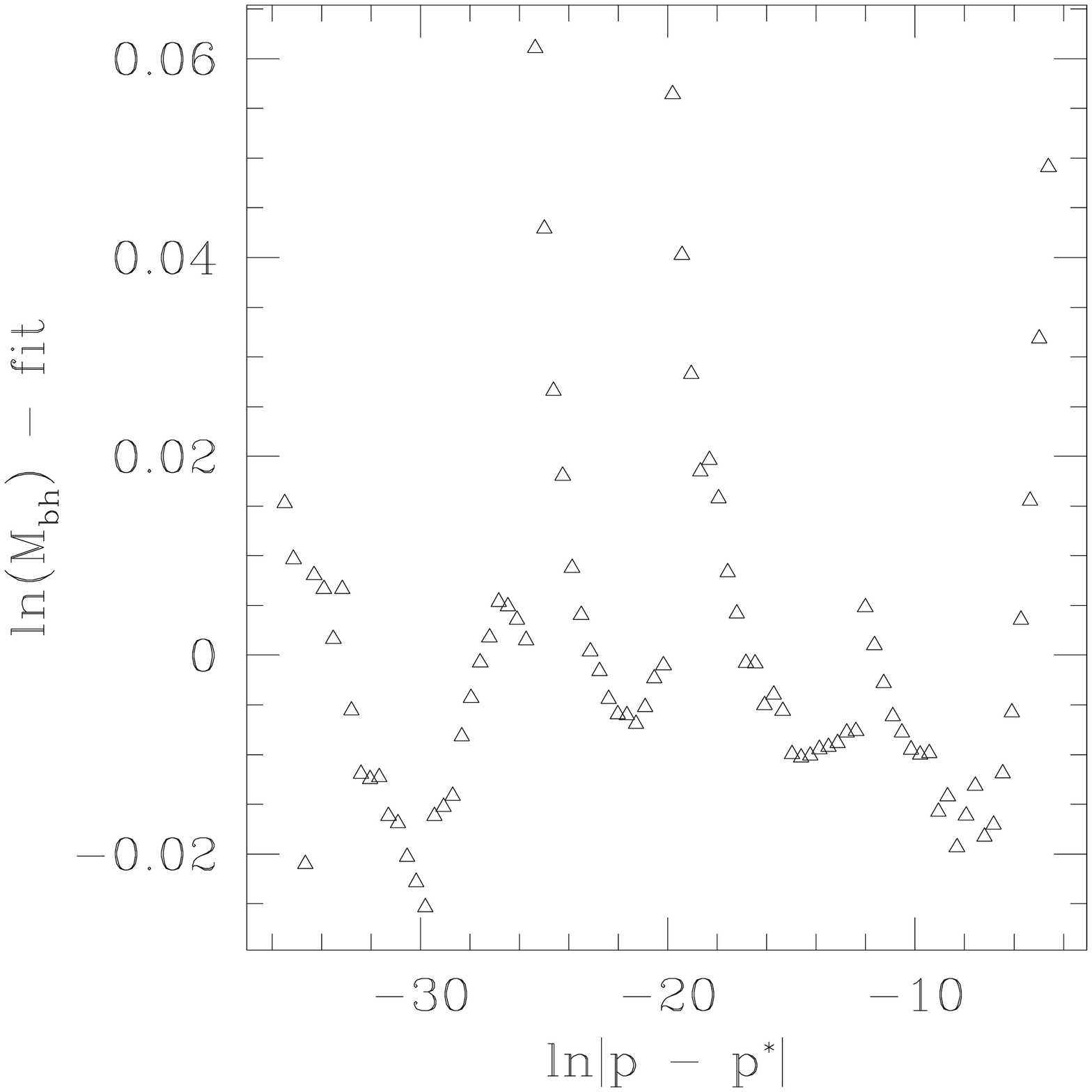}
\caption{Plot of the residuals of the data shown in  Fig.~\ref{bhplot} 
with respect to the computed linear fit. 
The absence of regular oscillations 
indicates that the geometry is not discretely self-similar.    
There are, however, features in the plot, notably the ``spikes'',  that can be 
explained as follows.
The matter fields have a discretely self-similar nature, but combine 
to produce a continuously self-similar geometry.  Truncation error 
effects, combined with the fact that our determination of black hole 
mass is not precise, result in a residual imprint of the DSS nature 
of the Dirac fields in the plot.
In fact the spikes in this plot occur as the amplitude of the Dirac 
fields reach an extremum near the black hole, periodically in 
$\ln | p - p^\star|$.}
\label{bherr}
\end{figure}

The data in Fig.~\ref{bhplot} does not oscillate
about the fit line since the geometry is continuously self-similar.  This
lack of regular oscillations about the fit line is shown more clearly in
Fig.~\ref{bherr}.

\section{Self-Similar Ansatz}
\label{sec:css}

Given the numerical evidence suggesting the existence of a self-similar 
solution at the black hole threshold, we proceed to an {\it ab initio}
computation
of the critical solution based on the application 
of a self-similar ansatz to our system.
The development here closely parallels the work 
done by Hirschmann 
and Eardley \cite{ode}, who considered the case of a massless, complex scalar field.

By definition,
a self-similar spacetime has a homothetic Killing vector, $\xi$, that obeys
\begin{equation}
\label{hkill1}
\mathcal{L}_{\xi} g_{\mu\nu} = 2g_{\mu\nu}
\end{equation}
where the factor $2$ is simply a matter of convention.  We want to define 
coordinates, $(\tau,x)$ that are adapted to this self-similar 
symmetry.  Specifically with $\tau$ adapted to the vector field $\xi^{\mu}$,
(\ref{hkill1}) can be written as:
\begin{equation}
\label{hkill2}
\frac{\partial}{\partial \tau} g_{\mu\nu}(\tau,x) = 2g_{\mu\nu}(\tau,x).
\end{equation}
Performing a separation of variables on the metric tensor and then
solving (\ref{hkill2}) for the $\tau$-dependent part yields
\begin{equation}
\label{metsep}
g_{\mu\nu}(\tau,x) = e^{2\tau}\tilde{g}_{\mu\nu}(x),
\end{equation}
where $\tilde{g}_{\mu\nu}(x)$ is the part of the metric that depends
only on $x$.
The original coordinates $(t,r)$ are related to $(\tau,x)$ by:
\begin{equation}
\label{ctrans}
\tau = \ln\left|\frac{t^{\star}-t}{L}\right|,\hspace{0.3in}
x = \frac{r}{t^{\star}-t}.
\end{equation}

The equations of interest will take the same form for any value of the constant factor, 
$L$, and without loss of generality, we subsequently take $L=1$.
The time $t^{\star}$ is the value of coordinate time to which the self-similar 
solution asymptotes as it propagates down to arbitrarily small scales, and, from the 
point of view of critical evolution, is a natural temporal origin for use in
describing the dynamics.  Further, as mentioned previously, in analyzing 
self-similar critical solutions, it is natural to adopt 
a parameterization of the $t$=constant surfaces, such that $t$ coincides with central
proper time.  We thus adopt such a normalization, and additionally adjust $t$ so that
$t^{\star}=0$.  

In these new coordinates, the metric~(\ref{metric}) becomes
\begin{equation}
\label{cssmetric}
ds^{2} = e^{2 \tau}\left[(- \alpha(x)^{2} + x^{2}a(x)^{2})d\tau^{2} 
+ 2xa(x)^{2}d\tau dx + a(x)^{2}dx^{2} + x^{2}d\theta^{2} +
x^{2}\sin^{2}\theta d\phi^{2}\right].
\end{equation}
We note that the $\tau$ coordinate is timelike, and, as can be verified by comparing the 
right hand sides of (\ref{metsep}) and (\ref{cssmetric}), that the functions, 
$\alpha$ and $a$ are functions of the spacelike coordinate, $x$, alone.

In $(\tau,x)$ coordinates, the spinors (\ref{spinor+}) and (\ref{spinor-})
become
\begin{equation}
\label{css_spinors}
\psi^{+} = \frac{e^{-\tau}}{2\sqrt{\pi}}\frac{e^{i\phi/2}}{x\sqrt{a(x)}}\left(
\begin{array}{c}
F(\tau,x)\sin(\theta/2)\\
G(\tau,x)\cos(\theta/2)\\
F(\tau,x)\sin(\theta/2)\\
G(\tau,x)\cos(\theta/2)
\end{array}
\right)
\end{equation}
\begin{equation}
\psi^{-} = \frac{e^{-\tau}}{2\sqrt{\pi}}\frac{e^{-i\phi/2}}{x\sqrt{a(x)}}\left(
\begin{array}{c}
  F(\tau,x)\cos(\theta/2)\\
- G(\tau,x)\sin(\theta/2)\\
  F(\tau,x)\cos(\theta/2)\\
- G(\tau,x)\sin(\theta/2)
\end{array}
\right),
\end{equation}
where these expressions 
were found by transforming the $(t,r)$ parts of (\ref{spinor+}) and 
(\ref{spinor-}) as scalars.  

In order to find spinor components that are only functions
of $x$, we require knowledge of the $\tau$ dependence of our field quantities.
This is determined by performing the coordinate transformations on the
equations of motion (\ref{eom1})-(\ref{eom4}) and the geometric equations
(\ref{hamcon}) and (\ref{lapse}), and then ascertaining what $\tau$ dependence
is needed in 
$F$ and $G$ to produce a set of $\tau$ independent ODEs.
A suitable ansatz is
\begin{equation}
F(\tau,x) = e^{\tau / 2}e^{i\omega \tau}x(P_{1}(x) + i P_{2}(x))
\label{Fansatz}
\end{equation}
\begin{equation}
G(\tau,x) = e^{\tau / 2}e^{i\omega \tau}x(Q_{1}(x) + i Q_{2}(x)).
\label{Gansatz}
\end{equation}
where the $\exp(i\omega t)$ terms reflect the fact that, as the PDE solutions 
have revealed, we expect the matter fields to exhibit discrete self-similarity.  Note that $\omega$ as defined here corresponds 
to $2\pi/\Delta$ (see \cite{ode})
where $\Delta$ is the echoing exponent originally defined in \cite{crit}.
Additionally, the extra factor of $x$ is introduced to cast the resulting 
equations in a more convenient form.  Inserting this ansatz into 
(\ref{eom1})-(\ref{eom4}), (\ref{hamcon}) and (\ref{lapse}), we find
\begin{equation}
\label{css1}
P_{1}^{\prime} = \frac{1}{x+\alpha/a}\left[- \frac{1}{2}P_{1}
- \omega P_{2} -
  \frac{1}{2}\frac{\alpha}{a}\left(\frac{a^{2}+1}{x}\right)P_{1}
+ 2\alpha P_{1}(P_{1}Q_{1} + P_{2}Q_{2}) + \frac{\alpha}{x}Q_{2}\right]
\end{equation}
\begin{equation}
P_{2}^{\prime} = \frac{1}{x+\alpha/a}\left[- \frac{1}{2}P_{2}
+ \omega P_{1} -
\frac{1}{2}\frac{\alpha}{a}\left(\frac{a^{2}+1}{x}\right)P_{2}
+ 2\alpha P_{2}(P_{1}Q_{1} + P_{2}Q_{2}) - \frac{\alpha}{x}Q_{1}\right]
\end{equation}
\begin{equation}
Q_{1}^{\prime} = \frac{1}{x-\alpha/a}\left[- \frac{1}{2}Q_{1}
- \omega Q_{2} +
  \frac{1}{2}\frac{\alpha}{a}\left(\frac{a^{2}+1}{x}\right)Q_{1}
- 2\alpha Q_{1}(P_{1}Q_{1} + P_{2}Q_{2}) + \frac{\alpha}{x}P_{2}\right]
\end{equation}
\begin{equation}
\label{css4}
Q_{2}^{\prime} = \frac{1}{x-\alpha/a}\left[- \frac{1}{2}Q_{2}
+ \omega Q_{1} +
\frac{1}{2}\frac{\alpha}{a}\left(\frac{a^{2}+1}{x}\right)Q_{2}
- 2\alpha Q_{2}(P_{1}Q_{1} + P_{2}Q_{2}) - \frac{\alpha}{x}P_{1}\right]
\end{equation}
\begin{equation}
\label{css_a}
\frac{a^{\prime}}{a} = \frac{1-a^{2}}{2x} + 2x\left( P_{1}P_{2}^{\prime}
- P_{1}^{\prime}P_{2} +  Q_{1}^{\prime}Q_{2} - Q_{1}Q_{2}^{\prime}
  \right)
+ 4a\left(P_{1}Q_{1} + P_{2}Q_{2}\right)
\end{equation}
\begin{equation}
\label{css_alpha}
\frac{\alpha^{\prime}}{\alpha} = \frac{a^{2}-1}{2x}
+ 2x\left( P_{1}P_{2}^{\prime}
- P_{1}^{\prime}P_{2} +  Q_{1}^{\prime}Q_{2} - Q_{1}Q_{2}^{\prime}
  \right).
\end{equation}
We note that we can cast this system into a canonical form suitable 
for numerical integration, by using (\ref{css1})-(\ref{css4}) 
to eliminate the derivatives of $P$ and $Q$ that
appear in the
right hand side of the equations for $a$ (\ref{css_a}) and $\alpha$
(\ref{css_alpha}).

\section{Numerical Solution of the Self-Similar ODEs}
\label{sec:shootres}

Having rewritten our equations in a coordinate system adapted to self-similar
symmetry, we can now solve the resulting ODEs to determine 
what we expect will be the CSS critical solution seen at the 
black hole threshold in the EMD model. 
Following Hirschmann and Eardley~\cite{ode}, we use a multi-parameter 
shooting method to integrate the equations subject to regularity and 
analyticity conditions.

We first observe that the system~(\ref{css1})-(\ref{css_alpha}) has 
singularities at $x = 0$ and at $x = x_{2} = \alpha/a$ (the similarity horizon).  
Of the infinitely many solutions to the ODEs, we seek one that is 
analytic at both of these points, as the CSS solution found via solution
of the PDEs has this property.
Our unknown problem parameters (``shooting'' parameters) include 
the values of some of the fields at the origin, the value of $\omega$ appearing in 
the ansatz~(\ref{Fansatz})-(\ref{Gansatz}), as well as the 
value of $x_{2}$ (the position of the similarity horizon).
Following Eardley and Hirschmann we 
shoot outwards from $x = 0$ and inwards from $x_{2} = \alpha/a$, 
comparing 
solutions at some intermediate point $x_{1}$.  
This process is automated by starting from some initial
guess, then using Newton's method to determine the shooting 
parameters for subsequent iterations. In the Newton iteration, 
we use the square of the differences of the values
of the functions and their derivatives at $x_1$ (computed from the 
inwards and outwards integrations)
as the goodness-of-fit indicator,
which is driven to 0 as the iteration converges.

At $x = 0$ we have the following:
\[
P_{1}(0) = 0
\]
\[
P_{2}(0) = - Q_{0}
\]
\[
Q_{1}(0) = Q_{0}
\]                                           
\[
Q_{2}(0) = 0
\]                                           
\[
\alpha(0) = 1
\]                                           
\[
a(0) = 1.
\]                                           
Regularity at the origin gives $P_{1} = Q_{2}$ and $P_{2} = - Q_{1}$.
We use the global $U(1)$ invariance of our system 
(\ref{css1})-(\ref{css_alpha}) to set $P_{1} = 0$.  This leaves $Q_{0}$ as a 
shooting parameter.

As noted previously, the location of the similarity horizon (outer
boundary of the integration domain) $x_{2}$ is itself a shooting parameter,
and is the value $x$ where $\alpha(x)/a(x) = x$.
In the limit $x \rightarrow x_{2}$ we have the following:
\[
P_{1}(x_{2}) = \frac{1}{2 a^{2}}\left(- 4 \alpha Q_{2}^{3}\omega 
               - 4\alpha Q_{1}^{2}Q_{2}\omega + 2 a \omega Q_{1}
               + Q_{2} a^{3}\right)
\]
\[
P_{2}(x_{2}) = \frac{1}{2 a^{2}}\left(4 \alpha Q_{1}Q_{2}^{2}\omega
               + 4\alpha Q_{1}^{3}\omega + 2 a \omega Q_{2}
               - Q_{1} a^{3}\right)
\]
\[
Q_{1}(x_{2}) = Q_{1}
\]
\[
Q_{2}(x_{2}) = Q_{2}
\]
\[
\alpha(x_{2}) = x_{2} a
\]
\[
a(x_{2}) = a.
\]
The shooting parameters at the outer boundary are: $x_{2}$, $Q_{1}(x_{2})$, 
$Q_{2}(x_{2})$, and $a(x_{2})$.  The final shooting parameter is the frequency, 
$\omega$, for a total of six undetermined parameters.

We find an approximate solution given by
\[
x_{2}  = 5.6740230 \pm 0.0000004
\]
\[
\omega = 4.698839  \pm 0.000001
\]
\[
Q_{1}(0) = 0.747912623 \pm 0.000000006
\]
\[
Q_{1}(x_{2}) = 0.00151341532 \pm 0.00000000007
\]
\[
Q_{2}(x_{2}) = 0.01103266083 \pm 0.00000000005
\]
\[
a(x_{2}) = 1.1183631604 \pm 0.0000000009.
\]                                     
where the quoted uncertainty was estimated by solving the system for
many different values of $x_{1}$ and observing the changes in the
shooting parameters. 

\section{Comparison of PDE/ODE Solutions}
\label{sec:com}

In this section we compare the solution computed from the self-similar 
ansatz, as just described, to the near-critical solutions calculated 
from the full PDEs in the $(t,r)$ coordinate system.  The ODE solution is
the theoretically predicted self-similar solution while the
PDE solution can be thought of as collected data.  For this comparison, 
we used data from the Gaussian family.  The idea is to treat
the ODE solution as the model function and fit it to the PDE data.
We perform the fit ``simultaneously'' at all times by working with the 
functions as $2$-dimensional solution surfaces in $t$ and $r$.  
This process is automated by starting from some initial
guess for the fitting parameters, then using Newton's method to determine 
these parameters for subsequent iterations.  In the Newton iteration,
we use the least squares of the two solutions:
\begin{equation}
\sum^{N}_{j=1} \left(u^{ODE}_{j} - u^{PDE}_{j} \right)^{2}
\end{equation}
as the goodness-of-fit indicator, which is driven towards 0 as the iteration 
converges.  When this happens, we compute the $l_{2}$-norm of the difference 
of the two solutions:
\begin{equation}
||u||_{2} = \left(\frac{1}{N} \sum^{N}_{j=1} \left(u^{ODE}_{j} -
u^{PDE}_{j} \right)^{2} \right)^{1/2}
\end{equation}
as an error estimate.  In computing both the least squares and the
$l_{2}$-norm, the solution to the ODEs and the solution to the PDEs are 
treated as $N$-element vectors where $N$ is the total number of grid 
points on the $2$-dimensional grid in $t$ and $r$.

\begin{figure}[ht]
\includegraphics[scale=0.6]{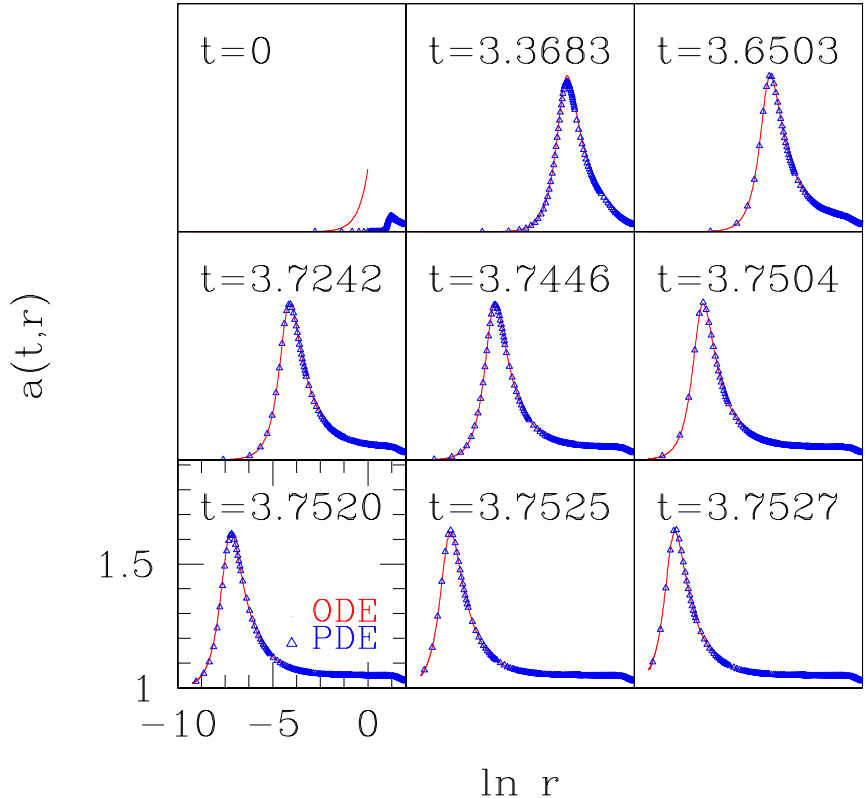}
\caption{Evolution of the metric variable, $a(t,r)$, for a slightly 
supercritical evolution,
overlayed with the solution to the ODEs.  
The frames are output
logarithmically in central proper time, as measured from $t^\star$.
The function's peak reaches a value 
of approximately $1.6$ and remains there as the solution continuously repeats 
itself on ever smaller scales.}
\label{aplot}
\end{figure}

For the case of $a(t,r)$, we use $t^{\star}$ as the fitting parameter.
Once $t^{\star}$ is found, the $l_{2}$-norm of the difference of the 
two solutions is $0.00159$.  We display the results of this fitting 
process as a sequence of snapshots in time in Fig.~\ref{aplot}.

\begin{figure}[ht]
\includegraphics[scale=0.6]{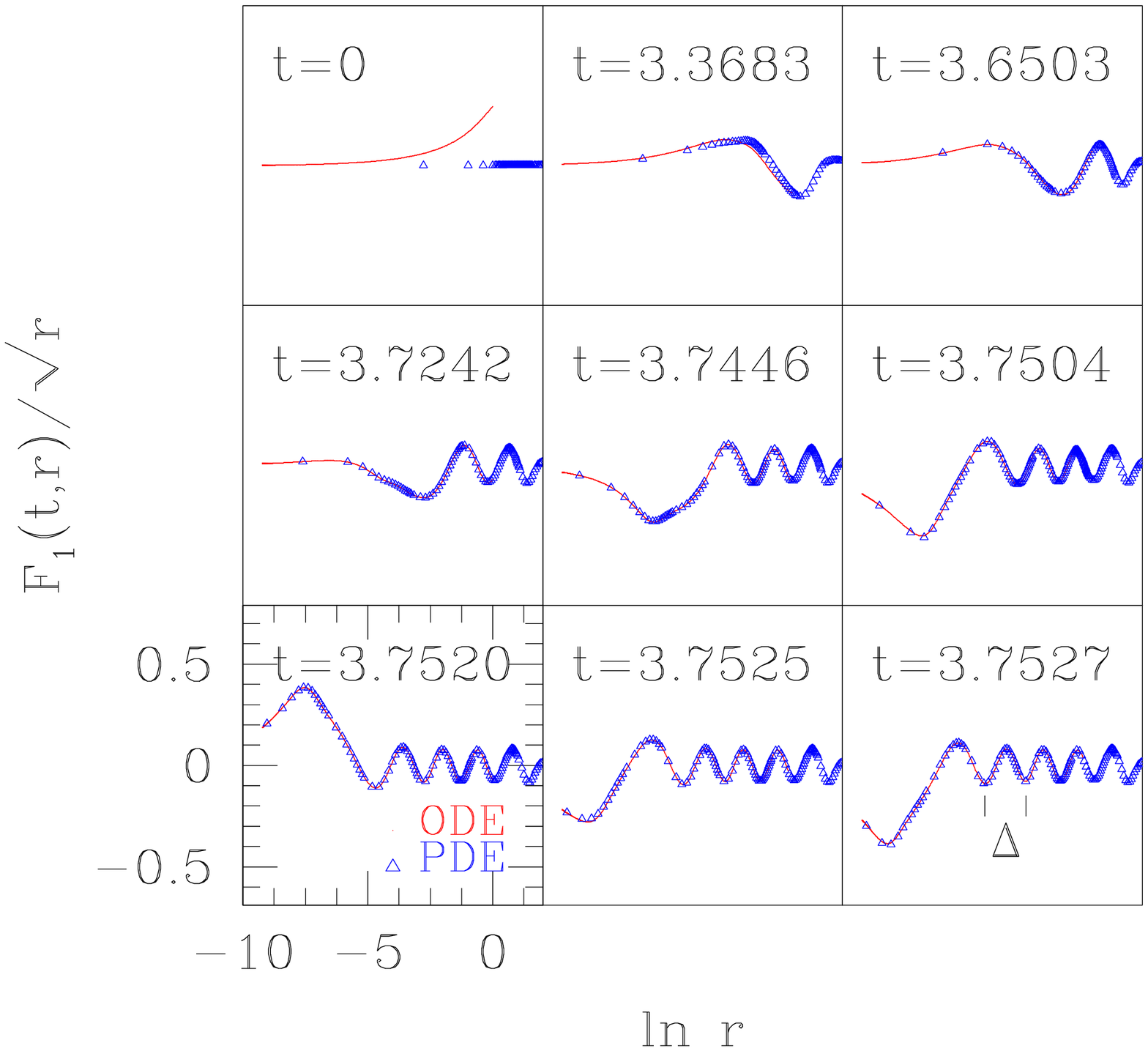}
\caption{Comparison of $F_{1}(t,r)$ for a slightly supercritical PDE 
evolution with the solution found from the self-similar ansatz.  
As in Fig.~\ref{aplot},
the frames are output logarithmically in central proper time, measured
from $t^\star$.  The 
function oscillates and discretely repeats itself on smaller and smaller 
scales.  The particular combination of $F_{1}(t,r)/\sqrt r$ is the natural scaling 
variable as can be deduced by noting that the Dirac fields have units of 
$($Length$)^{1/2}$, and as shown explicitly by the $\tau$ dependence of
(\ref{Fansatz}) and (\ref{Gansatz}).  The echoing exponent, 
$\Delta$, calculated via $\Delta = 2\pi / \omega$, where $\omega$ is 
computed from the ODE solution, is also shown.}
\label{F1plot}
\end{figure}

Comparing the Dirac fields is a little more involved since there are a
number of unspecified parameters and phases that must be determined.
From the ansatz (\ref{Fansatz})-(\ref{Gansatz}) we have:
\[
F_{1}(t,r) = \frac{rA_{1}}{(t^{\star} - t)^\frac{1}{2}}\left[
P_{1}(t,r)\cos(\omega \ln(t^{\star}-t)+\phi_{1}) - P_{2}(t,r)\sin(\omega 
\ln(t^{\star}-t)+\phi_{1})\right]
\]
\[
F_{2}(t,r) = \frac{rA_{2}}{(t^{\star} - t)^\frac{1}{2}}\left[
P_{1}(t,r)\sin(\omega \ln(t^{\star}-t)+\phi_{2}) + P_{2}(t,r)\cos(\omega 
\ln(t^{\star}-t)+\phi_{2})\right]
\]
\[
G_{1}(t,r) = \frac{rA_{2}}{(t^{\star} - t)^\frac{1}{2}}\left[
Q_{1}(t,r)\cos(\omega \ln(t^{\star}-t)+\phi_{2}) - Q_{2}(t,r)\sin(\omega 
\ln(t^{\star}-t)+\phi_{2})\right]
\]
\begin{equation}
G_{2}(t,r) = \frac{rA_{1}}{(t^{\star} - t)^\frac{1}{2}}\left[
Q_{1}(t,r)\sin(\omega \ln(t^{\star}-t)+\phi_{1}) + Q_{2}(t,r)\cos(\omega 
\ln(t^{\star}-t)+\phi_{1})\right].
\end{equation}
We note that $F_{1}$ and $G_{2}$ have the same phase, $\phi_{1}$, while the
pair $F_{2}$ and $G_{1}$ have the same phase $\phi_{2}$.  This is
expected from the coupling of (\ref{eom1})-(\ref{eom4}).
The equations of motion may be invariant under changes of these phases,
but (\ref{hamcon}) and (\ref{lapse}) are not.  In order to have the
entire system be invariant under changes in the phases, we must have:
\[
A_{1}A_{2} = \frac{1}{\cos(\phi_{1} - \phi_{2}) }
\]
We see that the amplitudes of the fields must change only if the {\it
relative} phase, $\phi_{1} - \phi_{2}$, changes.  We merely note this
fact for completeness but do not use it to reduce the number of fit
parameters.

The comparison of the fields as found from the PDEs and ODEs is carried out in much the same way as it
is done for the metric variable, $a(t,r)$.  The
goodness-of-fit is again defined to be the least squares of the two solutions
but this time, the parameter  $t^{\star}$ is kept fixed and the phase
$\phi_{j}$ and amplitude $A_{j}$ are used as fitting parameters 
($j = 1,2)$.  The $l_{2}$-norm of the difference of the solutions for $F_{1}$ 
is $0.000195$.  The $l_{2}$-norm of the difference of the solutions for $G_{1}$
is $0.00024$.  Fig.~\ref{F1plot} illustrates the results of this comparison of 
the ODE and PDE solutions for $F_{1}$.

\section{Conclusions }
\label{sec:concl}
We have investigated the spherically symmetric Einstein-massless-Dirac
system at the threshold of black hole formation.  We have found strong 
evidence for a
Type II critical solution, and an associated mass scaling law with 
a universal exponent $\gamma \sim 0.26$.
The solution exhibits continuous self-similarity in the
geometric variables and discrete self-similarity in the components of the Dirac
fields, the latter characterized by an echoing exponent $\Delta \sim 1.34$.  
Using a self-similar ansatz, we then reduced the equations of 
motion governing the model to a set of
ODEs whose solution, given appropriate regularity conditions, is in 
very good agreement with the critical solution 
obtained from the original PDEs.

\begin{acknowledgments}
We would like to thank Bill Unruh, Luis Lehner, and the rest of the members of 
the numerical relativity group at the University of British Columbia for 
useful discussions.  We 
would also like to thank Daniel A.~Steck and Douglas W.~Schaefer, who worked 
with us on investigating dynamic solutions of the massive Dirac problem that was the original inspiration for this work.  
That project was, in turn, inspired by a paper written by Finster, 
Smoller, and Yau \cite{fsy} in which solutions for a static, massive 
Einstein-Dirac field were presented.

We would like to acknowledge financial support from the National Science
Foundation grant PHY9722068 and from a grant from the Natural Sciences
and Engineering Research Council of Canada (NSERC).
This work was also supported in part by grant NSF-PHY-9800973 and by
the Horace Hearne Jr. Institute for Theoretical Physics.
\end{acknowledgments}

\end{document}